\documentclass[aps,preprint,superscriptaddress,showkeys]{revtex4-1}
\usepackage{graphicx}
\usepackage{amsmath}
\usepackage{amssymb}
\usepackage{longtable}
\usepackage{dcolumn}
\usepackage[sort&compress]{natbib}
\begin{document}

\title{Solvophobic and solvophilic contributions in the water-to-aqueous guanidinium chloride transfer free energy of model peptides}
\author{Dheeraj S.\ Tomar}
\affiliation{Chemical and Biomolecular Engineering, Johns Hopkins University, Baltimore, MD}\thanks{Current address: Pfizer, St.\ Louis, MO}
\author{Niral Ramesh}
\affiliation{Chemical and Biomolecular Engineering, Rice University, Houston, TX} 
% \author{Val{\'e}ry Weber}
% \affiliation{IBM Research, Zurich, CH-8803 R\"{u}schlikon, Switzerland}
\author{D. Asthagiri}\thanks{Email: dna6@rice.edu}
\affiliation{Chemical and Biomolecular Engineering, Rice University, Houston, TX} 
\date{\today}

\begin{abstract}
We study the solvation free energy of two different conformations (helix and extended) of two different peptides (deca-alanine and deca-glycine)
in two different solvents (water and aqueous guanidinium chloride, GdmCl).  The free energies are obtained using the quasichemical organization of the potential distribution theorem, an approach that naturally provides the repulsive (solvophobic or cavity) and attractive (solvophilic) contributions to solvation. The solvophilic contribution is further parsed into a chemistry contribution arising from solute interaction with the solvent in the first solvation shell and a long-range contribution arising from non-specific interactions between the solute and the solvent beyond the first solvation shell. The cavity contribution is obtained for two different envelopes,  $\Sigma_{SE}$ which theory identifies as the solvent excluded volume and a larger envelope ($\Sigma_G$) beyond which solute-solvent interactions are Gaussian. For both envelopes,  the cavity contribution in water is proportional to the surface area of the envelope. The same does not hold for GdmCl(aq), revealing limitations of using molecular area to assess solvation energetics, especially in mixed solvents. The $\Sigma_G$-cavity contribution predicts that GdmCl(aq) should favor the more compact state, contrary to the role of GdmCl in unfolding proteins.   The chemistry contribution attenuates this effect, but still the net local (chemistry plus $\Sigma_G$-packing) contribution is inadequate in capturing the role of GdmCl.  With the inclusion of the long-range contribution, which is dominated by van~der~Waals interaction, aqueous GdmCl favors the extended conformation over the compact conformation. Our finding emphasizes the importance of  weak, but attractive, long-range dispersion interactions in protein solution thermodynamics. 
\end{abstract}

% The $\Sigma_{SE}$-cavity contribution is also more unfavorable in GdmCl than in water, but the relative balance between the conformers varies with the type of peptide.
\maketitle

\section{Introduction}

The thermodynamics of protein-solvent interactions are of principal interest in determining the physical contributions to protein stability. In the quest to understand protein-solvent interactions, solvent additives termed cosolutes to the protein have often been used to perturb the solvation thermodynamics of the protein. By modeling the perturbation in terms of the effect of the cosolute on individual groups of the protein, insights have been sought into the molecular determinants of protein stability (for example, see \cite{tanford:apc70,pace:jbc74,Auton:bc11}). Here, based on modern developments in the calculation of protein solvation thermodynamics from computer simulations \cite{Weber:jctc12,tomar:bj2013,tomar:jpcb14,tomar:jpcb16,asthagiri:gly17}, we study the role of the 
cosolute guanidinium chloride, GdmCl, a protein denaturant, in modulating the solvation thermodynamics of model proteins. A distinguishing aspect of our work is that we naturally obtain insights into the solvophilic (hydrophilic) and solvophobic (hydrophobic) contributions to solvation, quantities of fundamental importance in protein solution thermodynamics, while also avoiding the approximations inherent in the traditional group-additive description of protein solvation.

The metric of interest in characterizing the role of the cosolute in modulating the (folded) F$\rightleftharpoons$U (unfolded) transition of the peptide
 is
\begin{eqnarray}
\Delta G & = &  [G_{U,{(c)}} - G_{F,{(c)}}] - [G_{U,{(w)}} - G_{F,{(w)}}] \nonumber \\
              & = & [G_{U,{(c)}} - G_{U,{(w)}}] - [G_{F,{(c)}} - G_{F,{(w)}}]
\label{eq:dg}        
\end{eqnarray}
where $G_{\alpha,{(x)}} $ is the free energy of the $\alpha$-conformational state in solvent $x$, where $x$ is either water ($w$) or the cosolute ($c$) solution. (In general, the $\alpha$-conformational state can comprise an ensemble of conformations.)  For denaturants $\Delta G < 0$. 

Formally, $G_{\alpha,{(x)}} = G_\alpha^{(vac)} + \mu^{\rm ex}_{\alpha,(x)}$,  where $G_\alpha^{(vac)}$ is the free energy in the absence of the solvent and $\mu^{\rm ex}_{\alpha,(x)}$ is the solvation free energy of the $\alpha$-conformational state in the solvent $x$.  $G_\alpha^{(vac)}$ subsumes the intramolecular effects in the $\alpha$-conformational state, but since this term is independent of the solvent, we find that 
\begin{eqnarray}
\Delta G & = &  [\mu^{\rm ex}_{U,(c)} - \mu^{\rm ex}_{U,(w)}] - [\mu^{\rm ex}_{F,(c)} - \mu^{\rm ex}_{F,(w)}] \nonumber \\ 
& = & \Delta \mu^{\rm tr}_{U} - \Delta \mu^{\rm tr}_{F} \, ,
\label{eq:dg1}        
\end{eqnarray}
where the quantities $\Delta \mu^{\rm tr}_{F}$ and $\Delta \mu^{\rm tr}_{U}$ are the water-to-cosolute solution transfer free energies of, respectively, the folded and unfolded conformational states. 

Following Tanford's seminal studies \cite{tanford:apc70,tanford:62,Tanford:1964}, it is common to set
\begin{eqnarray}
\Delta \mu^{\rm tr}_{\alpha} =  \sum_i f_{\alpha, i} \Delta g^{{\rm tr}}_{i} \, ,
\label{eq:additive}
\end{eqnarray}
where $i$ denotes the protein group, $\Delta g^{{\rm tr}}_i$ is the transfer free energy of the group $i$ at the given concentration of the cosolute, and $f_{\alpha,i}$ is a measure of the fractional solvent exposure of the group $i$ in the conformational state $\alpha$. The peptide group and the individual residues are typically the groups used in such a group-additive description. 

Eq.~\ref{eq:additive} has anchored efforts that seek a molecular-scale rationalization of experimental data (for example, see \cite{tanford:apc70,pace:jbc74,Auton:bc11,tanford:62,Tanford:1964}). However, emerging computational studies reveal important limitations of Eq.~\ref{eq:additive}. First, as discussed extensively in our earlier works \cite{tomar:bj2013,tomar:jpcb14}, there are nontrivial correlations between the individual groups comprising the protein that limit the validity of Eq.~\ref{eq:additive}, a conclusion that has been reached by other researchers as well \cite{helms:2005fw,baldwin:proteins06,Boresch:jpcb09}.  (Dill \cite{Dill:1997tg} has argued that for Eq.~\ref{eq:additive} to be useful in making predictions, systematic errors should be absent and the random error in the individual transfer free energies must be small; but as our earlier work shows \cite{tomar:bj2013,tomar:jpcb14}, the inter-group correlations easily outweigh the random errors.)  The second issue relates to the estimation of the solvent exposure.  The scaling of the $\Delta g^{\rm tr}_i$ by $f_{\alpha, i}$ is intuitively reasonable, but simultaneously mapping  both solvophobic and solvophilic aspects of solvation using a single local property of the solute, namely its molecular surface, implicitly assumes the same length scale dependence for both contributions.  Moreover, the molecular surface itself is usually calculated as the surface that is accessible to a water-sized probe (for example, see \cite{Auton:bc11,Courtenay2000}), whereas the solvent comprises multiple components of different sizes and shapes that interact with the solute. 

Computer simulations can, in principle, avoid some of the approximations noted above, provided the underlying model of interactions are reasonable and the simulations are well-crafted.  However, except for a few studies \cite{Hu:proteins2010,Hu:2010b,pettitt:jacs11,tomar:bj2013}, cosolute effect on protein transfer free energies have not been sought in computer simulations, likely because of the theoretical and computational intricacy in calculating free energies for solvated proteins. Simulation studies have usually focused on the free energy change along a reaction coordinate. Such potential of mean force (PMF) calculations provide insights into the role of the cosolute on protein ensembles (for example, see \cite{OBrien:jacs07,garde:2010gdmcl,jungwirth:jpcb11,pappu:jacs15}) and have an important advantage in avoiding approximations about the structures involved. But by the very nature of the sampling procedure, since the structures  necessarily differ in water and the cosolute solution, a clear comparison of solvent effects on the same structure and thus an estimation of transfer free energy is not possible.

In previous works we have presented an approach that makes possible the facile calculation of free energies of hydration (solvation) of polypeptides and proteins in all-atom simulations \cite{Weber:jctc12,tomar:bj2013,tomar:jpcb14,tomar:jpcb16,asthagiri:gly17}.  Thus all the $\mu^{\rm ex}$ factors in Eq.~\ref{eq:dg1} can be obtained. This approach also provides a direct quantification of the hydrophilic (solvophilic) and hydrophobic (solvophobic) contributions to hydration (solvation). The potential distribution approach also allows, in principle, 
the calculation of the solvation free energy of an ensemble of structures, although in practice the computational challenge is still daunting. Thus the 
work developed here offers the prospect of complementing and enhancing the interpretation of both the transfer-free-energy-based analysis of experimental data and the PMF-based simulation approaches. 

For the specific case of aqueous GdmCl, the present work shows that if solvophobicity were the dominant factor in protein stability, aqueous GdmCl ought to favor the compact protein structure, contrary to experimental observations. We find that solvophilicity plays an important role in favoring the extended peptide structure over the more compact structure. Analysis of the solvophilic contributions shows that turning on electrostatic interactions tends to temper the effect of the cosolute, but crucially, long-range dispersion interactions that are often ignored in discussions of solvent effects play an important  role in the solution thermodynamics. 

\section{Theory}

In calculating $\mu^{\rm ex}_{\alpha,x}$ above, in general, one must consider an ensemble of conformations belonging to the defined
conformational state.  As we showed earlier, from a multi-state generalization \cite{asthagiri:gly17} of the potential distribution theorem \cite{lrp:cpms,lrp:book}, 
for a given solvent
\begin{eqnarray}
\beta \mu^{\rm ex}_{\alpha} & = &  \ln  \int_{\xi\, \in\, \alpha} e^{\beta \mu^{\rm ex} (\xi)} P(\xi) d \xi \nonumber \\
 & = & \beta \mu^{\rm ex} (\xi^\star) + \ln  \int_{\xi\, \in\, \alpha} e^{\beta [\mu^{\rm ex} (\xi)-\mu^{\rm ex}(\xi^\star)]} P(\xi) d \xi
\label{eq:multistate}
\end{eqnarray}
where $\xi$ is an order parameter characterizing the conformation and $\mu^{\rm ex}(\xi)$ is the excess chemical potential of that conformation. $\xi^\star$ is the order parameter for the least solvated structure, i.e.\ the structure with the highest
solvation free energy.  $P(\xi) d\xi$ is probability of finding a conformation in the range $[\xi,\xi+d\xi]$ and the integration is over the ensemble of conformations belonging to state $i$. As usual,  $\beta = 1/k_{\rm B}T$, with $k_{\rm B}$ the Boltzmann constant and $T$ the temperature. 

Characterizing $P(\xi)$ completely and calculating the corresponding $\mu^{\rm ex}(\xi)$ using all-atom simulations is a formidable challenge, but we can make further progress by making reasonable approximations. First note that $\mu^{\rm ex}(\xi)$ values are typically negative. Further, Eq.~\ref{eq:multistate} makes it clear that $\mu^{\rm ex}_{i} \leq \mu^{\rm ex}(\xi^\star)$.  As we found previously from
an analysis of several distinct conformations of a polyglycine chain \cite{asthagiri:gly17}, the solvation free energies tend to decrease rather rapidly relative to the change in $-k_{\rm B}T\ln P(\xi)$. Thus, the least negative free energy itself is a reasonable initial approximation of the solvation free energy of the ensemble. 

With the above framework as a basis, we consider two limiting structures. For the folded conformer, we choose an ideal (Ala)$_{10}$ helix. For the unfolded conformer, we choose the extended (Ala)$_{10}$ conformer with the least negative hydration free energy. This conformer was labelled $C_0$ in our previous study \cite{tomar:jpcb16}. For each of these conformers, we also construct the polyglycine analogues. Exploratory adaptive bias force \cite{abf1,abf2} calculations of the potential of mean force with the end-to-end distance as a reaction coordinate suggests that the least hydrated structure in water will also be the least solvated structure in aqueous GdmCl.

\subsection{Solvation free energy --- $\mu^{\rm ex}$}
As discussed previously  \cite{weber:jcp11,Weber:jctc12,tomar:bj2013,tomar:jpcb14,tomar:jpcb16,asthagiri:gly17}, we regularize the statistical problem of calculating the excess chemical potential using the potential distribution theorem $\beta \mu^{\rm ex} =  \ln \langle e^{\beta \varepsilon}\rangle$ \cite{widom:jpc82,lrp:book}, where the averaging $\langle\ldots\rangle$ is over the solute-solvent binding energy ($\varepsilon$) distribution $P(\varepsilon)$, and $\mu^{\rm ex}$ is defined relative to the ideal gas at the same density and temperature.  
The regularization is achieved by introducing an  auxiliary variable, a field $\phi(\lambda; r)$ that moves the solvent away from the solute \cite{weber:jcp11,Weber:jctc12,tomar:bj2013,tomar:jpcb14,tomar:jpcb16,asthagiri:gly17}. The distance between the center of the field and the solvent molecule is $r$. For $r > \lambda$, $\phi=0$. Since the solvent interface is pushed away from the solute, the solute-solvent interaction in the presence of the field is tempered and the conditional distribution $P(\varepsilon|\phi)$ is better behaved than $P(\varepsilon)$. In practice, we adjust the range $\lambda$ such that $P(\varepsilon|\phi)$ is Gaussian, rendering the calculation of $\mu^{\rm ex}[P(\varepsilon|\phi)]$ statistically robust \cite{weber:jcp11,Weber:jctc12,tomar:bj2013,tomar:jpcb14,tomar:jpcb16,asthagiri:gly17}. 

\begin{figure}[h!]
\includegraphics[width=3.25in]{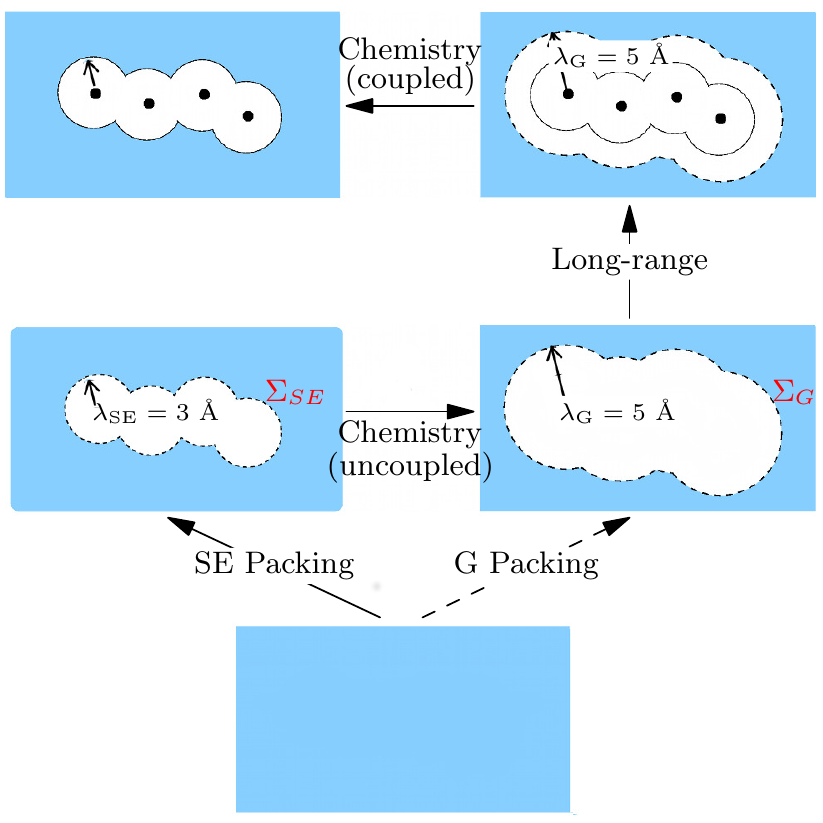}
\caption{Quasichemical organization of the excess chemical potential. $\Sigma_G$: the $\lambda_G = 5$~{\AA} envelope is conservatively the 
smallest region enclosing the solute for which the solute-solvent binding energy distribution $P(\varepsilon|\phi)$ is Gaussian. It approximately corresponds to the first solvation shell of the solute. $\Sigma_{SE}$: the $\lambda_{SE} \leq 3.0$~{\AA} envelope for which the chemistry contribution
is zero. Thus $\Sigma_{SE}$ encloses the volume excluded to the solvent.  Figure adapted with permission  from Tomar et~al., Journal of Physical Chemistry B, {\bf 120}, 69, 2016, Copyright (2016) American Chemical Society.}\label{fg:cycle}
\end{figure}

With the introduction of the field, we obtain \cite{weber:jcp11,Weber:jctc12,tomar:bj2013,tomar:jpcb14,tomar:jpcb16,asthagiri:gly17} the quasichemical organization of the potential distribution theorem
\begin{eqnarray}
\beta\mu^{\rm ex} = \underbrace{\ln x_0[\phi]}_{\rm chemistry}  + \underbrace{\beta\mu^{\rm ex}[P(\varepsilon|\phi)]}_{\rm long-range} \underbrace{- \ln p_0[\phi]}_{\rm packing}\, .
\label{eq:qc}
\end{eqnarray}
Fig.~\ref{fg:cycle} provides a schematic description of Eq.~\ref{eq:qc}. 
 $-k_{\rm B}T \ln x_0[\phi(\lambda)]$ is the free energy to apply the field in the presence of the solute; for $\lambda$ corresponding to the first solvation shell, the negative of this quantity is a measure of the gain in free energy when we allow local, specific solute-solvent interaction (Fig.~\ref{fg:cycle}) and is thus termed the chemistry contribution. $-k_{\rm B}T \ln p_0[\phi(\lambda)]$ is the free energy to apply the field to create a cavity in the solvent and is a measure of packing (or solvophobic) contributions to solvation. In the case of water, this contribution is a measure of hydrophobic contributions to hydration of the solute. 
$\beta\mu^{\rm ex}[P(\varepsilon|\phi)]$ is the contribution to the interaction free energy in the presence of the field and is a measure of the outer or long-range contribution to solvation. 

We apply the field about each heavy atom to carve the molecular shape in the liquid (Figure~\ref{fg:cycle}). For convenience we use the same $\lambda$ for each heavy atom; in practice, at the size-scale of the peptide, the distinction between the first hydration shell of N, C, and O heavy atoms is small enough that the choice of a single $\lambda$ is inconsequential. Further, we find that $\ln x_o \approx 0$  for $\lambda_{SE} \leq 3.0$~{\AA}, irrespective of the conformation of the peptide or the nature of the solvent (neat water or aqueous GdmCl).  This suggests that the space enclosed by $\Sigma_{SE}$ (Fig.~\ref{fg:cycle}) is excluded to the solvent and thereby provides a natural definition of the cavity to be used in examining solvophobic contributions to solvation. Thus we can recast Eq.~\ref{eq:qc} as 
\begin{eqnarray}
\beta \mu^{\rm ex} & = &  \underbrace{ \ln \left[x_0(\lambda_G)\frac{p_0(\lambda_{\rm SE})}{p_0(\lambda_G)}\right]}_{\rm revised\, chemistry} \underbrace{-\ln p_0(\lambda_{\rm SE})}_{\rm SE\, packing} +  \underbrace{\beta\mu^{\rm ex}[P(\varepsilon|\lambda_G)]}_{\rm long-range} \, .
\label{eq:qc1}
\end{eqnarray}
The revised chemistry contribution is a measure of the solute interaction with the solvent in the inner shell relative to the non-interacting solute. Thus it is a sensitive measure of the local contributions of attractive solute-solvent interactions to the free energy. The revised chemistry plus SE-packing contribution together gives the net local contribution to solvation. 

We finally note a subtlety involving the regularization approach. Throughout this work, we use a smooth-repulsive field in the regularization approach. In theoretical discussions of solvophobic effects, usually a hard-cavity is implied. The soft-cavity results can be easily corrected to get the hard-cavity results \cite{weber:jcp11}, but we do not pursue those calculations here since free energy differences involving the individual components (in Eqs.~\ref{eq:qc} or~\ref{eq:qc1}) are expected to be insensitive to this distinction. Thus, for brevity, we will refer to the soft-cavity packing result as the solvophobic contribution. 

\section{Methods}

The simulation method closely follows our earlier works \cite{tomar:bj2013,tomar:jpcb14,tomar:jpcb16}.  For the solvent components, 
we used the TIP3P model for water \cite{tip32,tip3mod} and the CGenFF parameter set for GdmCl.  For the peptide, we used version C31
of the CHARMM \cite{charmm} forcefield with correction (cmap) terms for dihedral angles. All the simulations comprise
3500 water molecules. For the aqueous GdmCl system, we used 500 GdmCl molecules, which gives a 5.2~M solution. 
Our choice of concentration was influenced by the choice made in experimental studies (for example, \cite{Tanford:1966,Rashid2005}). 

We also studied the transfer free energies by adapting the Kirkwood-Buff-forcefield model of GdmCl \cite{smith:gdcl04}. These results are qualitatively 
similar to those using CGenFF, and hence are not discussed further. 

The helix and extended C$_0$ conformations were as in our earlier study. The hydration results are also obtained from the earlier 
studies. For completeness, we briefly discuss the methods for the aqueous GdmCl system. All the simulations were performed using the NAMD \cite{namd,namd:2005} code at a temperature of 298~K using a Langevin thermostat and a pressure of 1 bar using a Langevin barostat. 

For solvation studies in aqueous GdmCl solution, we follow a procedure similar to that for water. To compute the chemistry or packing contribution, we need to build the field to its eventual range of $\lambda = 5$~{\AA}. To this end, we progressively apply the field, and for every unit {\AA} increment in the range,  we compute the work done in applying the field using a five-point Gauss-Legendre quadrature \cite{Hummer:jcp96}.  At each Gauss-point, the system was simulated for 1~ns and the (force) data from the last 0.5~ns used for analysis. (Excluding more data did not change the numerical value significantly, indicating good convergence.) Error analysis and error propagation was performed as before \cite{Weber:jctc12}: the standard error of the mean force was obtained using the Friedberg-Cameron algorithm \cite{friedberg:1970,allen:error} and in adding multiple quantities, the errors were propagated using standard variance-addition rules. 

The starting configuration for each $\lambda$ point is obtained from the ending configuration of the previous point in the chain of states. For the packing contribution, a total of 25 Gauss points span $\lambda \in [0,5]$.  For the chemistry contribution,
since solvent never enters $\lambda < 2.5$~{\AA}, we simulate $\lambda \in [2.5,5]$ for a total of 13 Gauss points.  For the water simulations, we had repeated the calculations four times and then averaged the results. With the benefit of hindsight and experience using the regularization approach \cite{tomar:jpcb16,asthagiri:gly17}, here we use a single well-converged simulation for GdmCl.  

The long-range contribution can be obtained using one of two forms
\begin{eqnarray}
\mu^{\rm ex}[P(\varepsilon|\lambda_G)] & = & \langle \varepsilon \rangle + \frac{\beta \sigma^2 }{2} \label{eq:inv} \\
\mu^{\rm ex}[P^{(0)}(\varepsilon|\lambda_G)] & = & \langle \varepsilon \rangle_0 - \frac{\beta \sigma^2 }{2} \label{eq:for}
\end{eqnarray}
where $P(\varepsilon|\lambda_G)$ is the distribution of binding energy with solute and solvent thermally coupled and $P^{(0}(\varepsilon|\lambda_G)$
is the distribution with solute and solvent thermally uncoupled. Both $P(\varepsilon|\lambda_G)$ and $P^{(0}(\varepsilon|\lambda_G)$ 
are assumed to be well-described by a Gaussian. Notice that
under the Gaussian description, $P(\varepsilon|\lambda_G)$ and $P^{(0)}(\varepsilon|\lambda_G)$ have the same variance ($\sigma^2$)
but the respective mean values, $\langle \varepsilon \rangle$ and $\langle \varepsilon \rangle_0$, differ.

For the aqueous GdmCl solution, for the coupled distribution, with $\lambda_G = 5$~{\AA} we perform a 2~ns long simulation, archiving frames every
200~fs. (The $\lambda_G = 5$~{\AA} calculations are started from the end-point of the chemistry calculation noted above.)  We then calculated the solute-solvent binding energy using the {\sc PairInteraction} module within NAMD.  Likewise, for the uncoupled distribution, using the end-point of the packing calculation, with $\lambda_G = 5$~{\AA} we perform a 2~ns long simulation, archiving frames every 200~fs. We then write out configurations adding the solute protein and then calculate $\varepsilon$ using the {\sc PairInteraction} module within NAMD. For the coupled and uncoupled distributions, we use the last 9500 configurations for calculating binding energies.

For the long-range contributions in water, earlier we had used Eq.~\ref{eq:for} and analyzed 10,000 snapshots repeating the calculations four times \cite{tomar:jpcb16}. With benefit of hindsight \cite{tomar:jpcb16,asthagiri:gly17}, we know the Gaussian description of $P(\varepsilon|\lambda_G)$ or $P^{(0)}(\varepsilon|\lambda_G)$ is robustly determined with far fewer points. Thus we split the 9500 energy values into blocks of 1900 values and using the resulting energy distribution obtain $\mu^{\rm ex}[P(\varepsilon|\lambda_G)]$ or $\mu^{\rm ex}[P^{(0)}(\varepsilon|\lambda_G)]$. The $\mu^{\rm ex}$ values obtained using Eqs.~\ref{eq:inv} and~\ref{eq:for} agree within the statistical uncertainty of either calculation, and for consistency with the earlier study we use only
the value based on Eq.~\ref{eq:for} in the analysis below. Note that when both $P(\varepsilon|\lambda_G)$ and $P^{(0}(\varepsilon|\lambda_G)$ are Gaussian, 
the long-range contribution can also be obtained as $0.5 (\langle \varepsilon \rangle +  \langle \varepsilon \rangle_0)$ \cite{beck:jcp08}.  For the
coil (helix) conformations, this estimate is within 3(9)\% of the value obtained using Eq.~\ref{eq:for} and within statistical uncertainty of the calculations, reaffirming the consistency between various approaches. 

The $P^{(0}(\varepsilon|\lambda_G)$ distribution can also be parsed into contributions from van~der~Waals and electrostatic interactions, with the 
reference being the cavity state. (Note that we cannot do the same decomposition of $P(\varepsilon|\lambda_G)$ without introducing non-intuitive
path-dependencies.) As shown below, to an excellent approximation, the individual van~der~Waals and electrostatic contributions are themselves Gaussian and uncorrelated from each other. 

To estimate the role of electrostatic interactions, we repeated the chemistry calculations with the partial charges on the peptides turned off ($Q=0$ peptides). 
The van~der~Waals component of $P^{(0}(\varepsilon|\lambda_G)$ also gave the long-range contribution for the $Q=0$ peptides. 

\section{Results}

\begin{figure}[h!]
\includegraphics[width=3.25in]{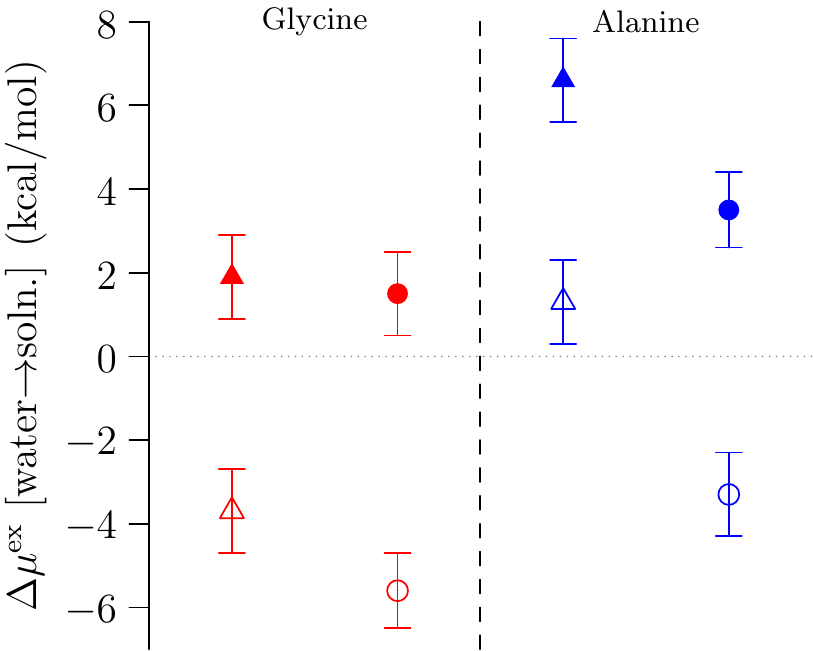}
\caption{Water-to-GdmCl(aq) transfer free energies. $\triangle$: helix; $\bigcirc$: $C_0$ coil conformer. The unfilled symbols correspond to $Q=0$ peptides.
Standard error of the mean is shown at $1\sigma$.}\label{fg:dmu}
\end{figure}

Figure~\ref{fg:dmu} shows the water-to-GdmCl(aq) transfer free energy for the different cases studied in this work. 
Notice that for the peptides considered here, the solvation in aqueous GdmCl is predicted to be less favorable than in water.  However, relative to the compact helical structure, the extended coil is better solvated in aqueous GdmCl.  Turning off the partial charges has the effect of lowering the transfer free energy 
relative to the corresponding peptide, indicating that peptide solvent electrostatic interactions temper the effect of GdmCl. 

\subsection{Packing contribution}
\begin{figure*}[h!]
\includegraphics[scale=1.]{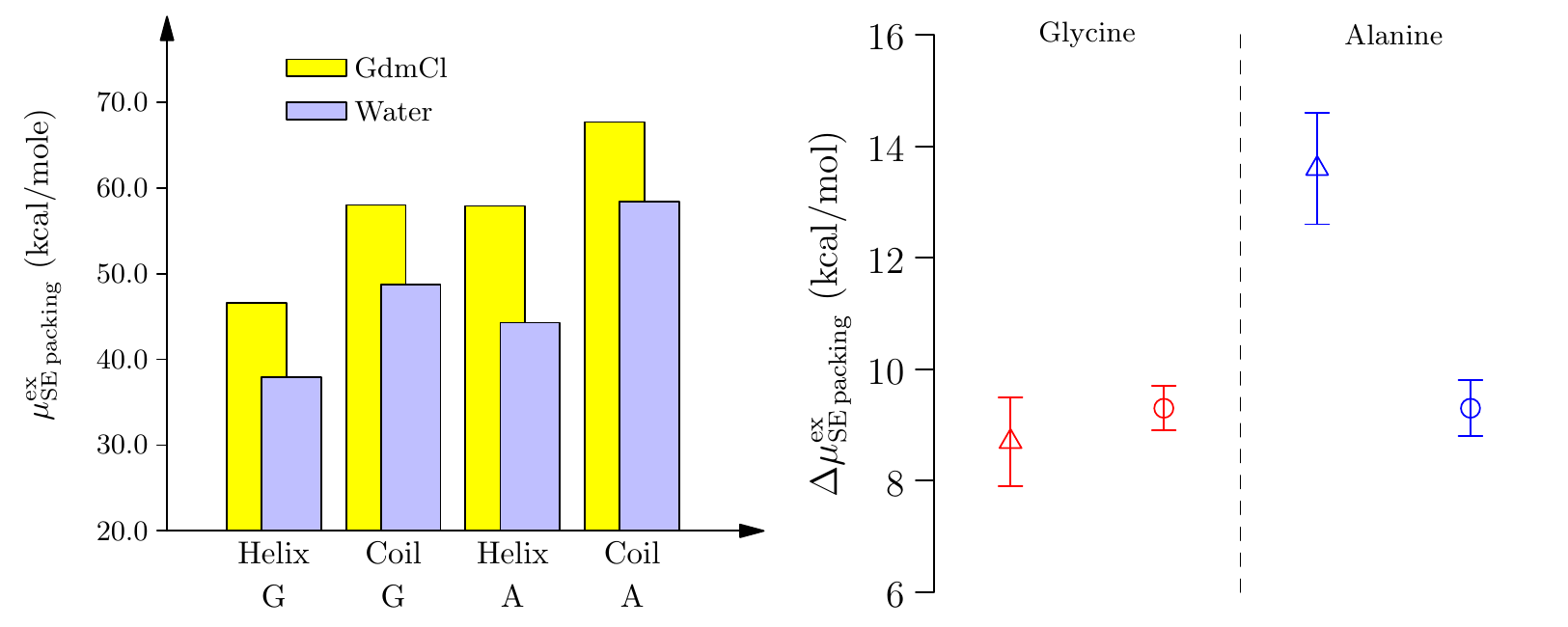}
\caption{\underline{Left panel}: Packing contribution for the $\Sigma_{SE}$ cavity (Fig.~\ref{fg:cycle}). G: glycine peptide; A: the alanine variant. \underline{Right panel}: Solvent-excluded packing contribution to the water-to-GdmCl(aq) transfer free energy. $\triangle$: helix; $\bigcirc$: $C_0$ coil conformer. Standard error of the mean is shown at $1\sigma$.}\label{fg:packing}
\end{figure*}

\begin{figure*}[h!]
\includegraphics[scale=1.]{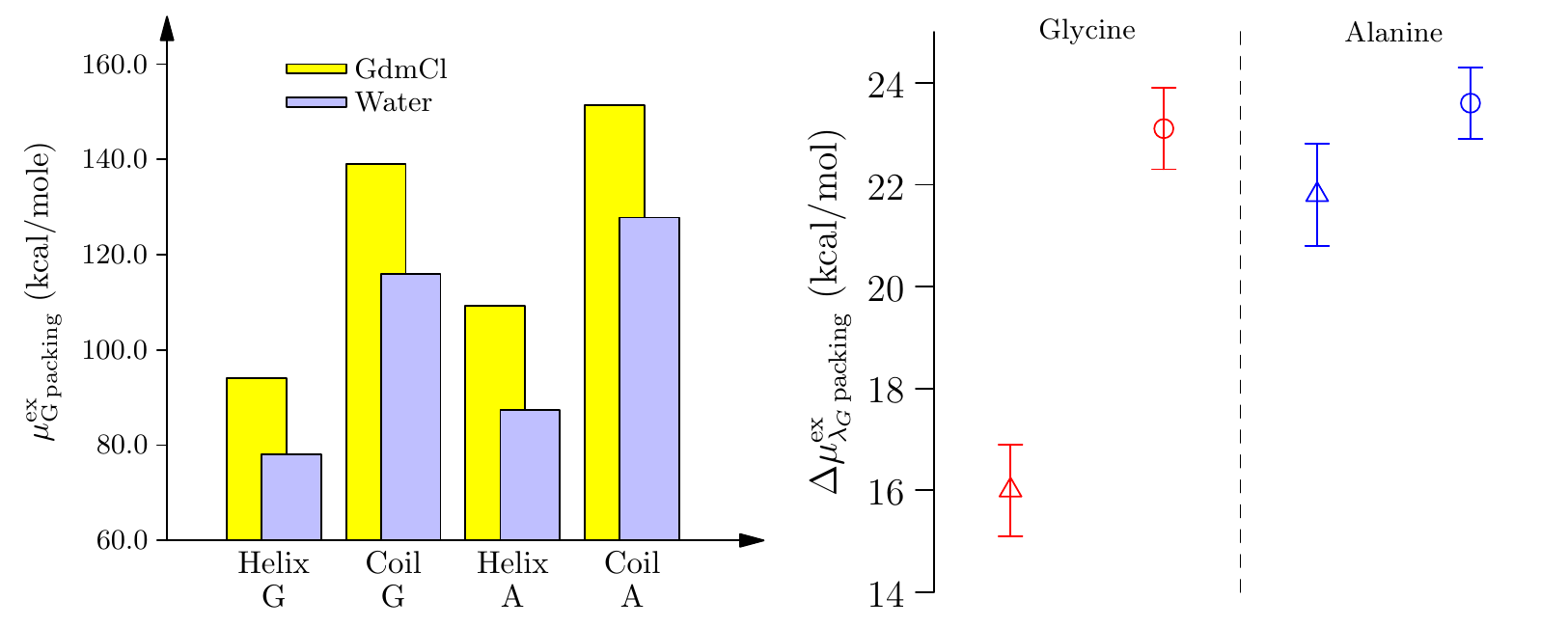}
\caption{\underline{Left panel}: Packing contribution for the $\Sigma_G$ cavity (Fig.~\ref{fg:cycle}). \underline{Right panel}: The corresponding
packing contribution to the water-to-GdmCl(aq) transfer free energy. Rest as in Figure~\ref{fg:packing}}\label{fg:Gpacking}
\end{figure*}

Figure~\ref{fg:packing} shows the packing contribution for the $\Sigma_{SE}$ cavity (Fig.~\ref{fg:cycle}). As expected, the packing contributions are
positive. Aqueous GdmCl increases the surface tension of the solution above that for pure water \cite{breslow:pnas90}; thus, perhaps not surprisingly, we find that the cost of creating a cavity in GdmCl is greater than the corresponding cost in liquid water.  
However, observe that the SE packing results do not evince a uniform trend: either GdmCl can stabilize the compact conformer, as suggested for the glycine peptide, or it can destabilize the compact conformer, as suggested for the alanine peptide (Fig.~\ref{fg:packing}, right panel).  Interestingly, for the larger $\Sigma_G$ envelope (Fig.~\ref{fg:Gpacking}), the packing contributions and the relative transfer free energies both 
conform to the intuitions based on surface tension. 

To rationalize the length-scale dependence of the packing contribution, we evaluate how the packing free energy scales with the area of the cavity. 
Table~\ref{tb:sasa} collects the ratio of the $\mu^{\rm ex}_{\rm SE\, packing}$ or $\mu^{\rm ex}_{\rm G\, packing}$ 
 to the surface area of the corresponding cavity, $\Sigma_{SE}$ or $\Sigma_{G}$, respectively. 

To calculate the surface area, all the heavy atoms are assigned a radius of either 3~{\AA} (for the $\Sigma_{G}$ envelope) or 5~{\AA} (for $\Sigma_{G}$ envelope).  Then using a probe of radius of 0.1~{\AA} we calculate the surface area using VMD. The small probe radius ensures that the 
calculated solvent accessible surface area is a good approximation to the molecular surface area, sometimes called the solvent
excluded surface area. Note that for $\Sigma_{SE}$, the molecular surface area is approximately the same as the usual solvent-accessible surface area (SASA) obtained using a solvent probe-radius of 1.4~{\AA} and reasonable atomic radii for heavy atoms. This agreement further reinforces
the consistency between what theory identifies as the solvent excluded volume, i.e.\ the volume enclosed by the $\Sigma_{SE}$ envelope, and SASA used
in biophysics. 
\begin{table*}
\caption{Ratio  $\gamma_{se} = \mu^{\rm ex}_{\rm SE\, packing} / Area(\Sigma_{SE})$ and $\gamma_{g} = \mu^{\rm ex}_{\rm G\, packing} / Area(\Sigma_{G})$, where $Area(\ldots)$ is the surface area of the corresponding envelope. Values are reported in kcal/mole-{\AA}$^2$.}\label{tb:sasa}
\begin{tabular}{l r r r r}
   & \multicolumn{2}{c}{\underline{Water}} & \multicolumn{2}{c}{\underline{GdmCl}} \\ 
Peptide &  $\gamma_{se}$ & $\gamma_{g}$ & $\gamma_{se}$ & $\gamma_{g}$ \\ \hline
Ala-Helix & 0.050 & 0.064 & 0.066 & 0.080 \\ 
Ala-Coil & 0.046 & 0.065 & 0.054 & 0.077 \\
Gly-Helix & 0.049 & 0.063 & 0.061 & 0.076 \\
Gly-Coil & 0.044 & 0.062 & 0.052 & 0.074 \\ \hline
\end{tabular}
\end{table*}

In water, for a given $\Sigma$, the corresponding scaling $\gamma$ (Table~\ref{tb:sasa}) is approximately constant across peptide types and protein conformers. This is especially so for $\Sigma_{G}$, a case where the smallest dimension of the cavity is several times the size of a water molecule.
(Please note that because the cavity is defined by a soft-repulsive wall, we cannot relate the scale factors to the precise value of the air-water surface tension.)
In GdmCl, the variation in $\gamma_{se}$ is larger than the corresponding variation for water. Note also that the scaling is different for the helix and the coil conformers. We suspect this is because the smallest dimension of the $\Sigma_{SE}$ cavity accommodating the coil, which is considerably smaller than the
smallest dimension of the $\Sigma_{SE}$ cavity for the helix, is comparable in size to the Gdm$^+$ cation. But just as we found for water, as the cavity size increases to $\Sigma_{G}$, the relative variation in $\gamma$ is reduced. 

The analysis above rationalizes the length-scale dependence of the packing contribution (Figs.~\ref{fg:packing} and~\ref{fg:Gpacking}). Importantly,
the analysis shows care is needed in using SASA-based approaches for obtaining solvation free energies, especially in mixed solvents.

\subsection{Chemistry contribution}

\begin{figure*}[h!]
\includegraphics[width=3.25in]{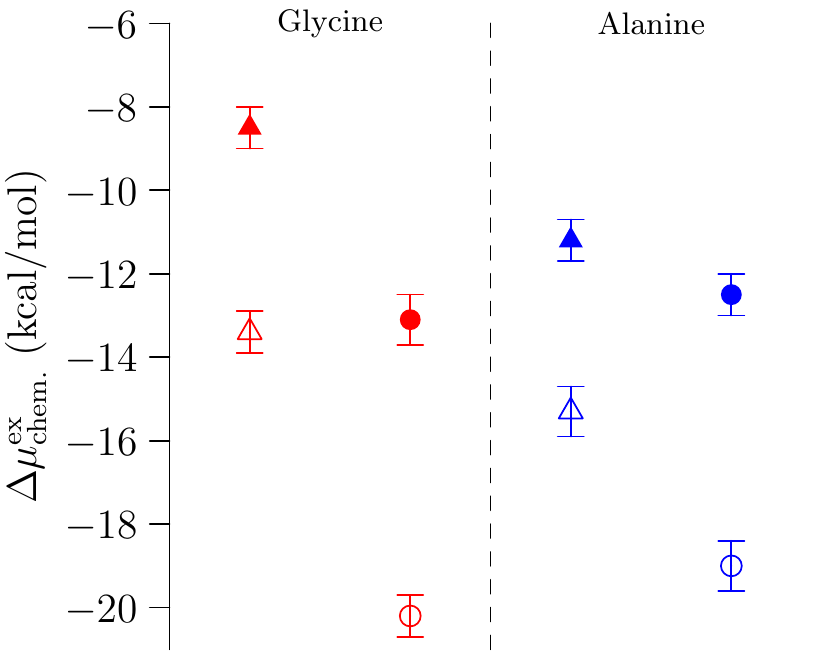}
\caption{The chemistry contribution (Eq~\ref{eq:qc}, Fig.~\ref{fg:cycle}) to the transfer free energy.   $\triangle$: helix; $\bigcirc$: $C_0$ coil conformer. The unfilled symbols correspond to $Q=0$ peptides.
Standard error of the mean is shown at $1\sigma$.}\label{fg:chemistry}
\end{figure*}

Figure~\ref{fg:chemistry} shows the total chemistry contribution to the transfer free energy. We find that relative to water the local interactions tend to stabilize the extended conformer to a greater extent than the compact conformer.  
Interestingly, the effect is more pronounced for the peptide lacking partial charges, reflecting the behavior seen in Fig.~\ref{fg:dmu}.

From a biophysical perspective, it is interesting to probe the nature of the local interaction between the cosolute and the protein. Of particular interest is
the possibility of hydrogen bonding between the cosolute and the peptide, something that can be probed in hydrogen exchange experiments \cite{englander:gdmcl09}. Here, for qualitative considerations, we analyze hydrogen bonding using a geometric criterion. From GdmCl, only the
carbonyl (C=O) groups of the protein can potentially hydrogen bond with the guanidinium N-H groups. We use a permissive condition to define a hydrogen bond:
if $r(O-N) \leq 3.5$~{\AA} and $\angle{OHN} \geq 145^\circ$ the bond is considered a hydrogen bond. In the deca-peptides with an acetylated N-terminus, there are 11 C=O groups that are available to hydrogen bond. By analyzing 2000 snapshots from a 2~ns simulation of the peptide in GdmCl, we find that on average less than one C=O group is hydrogen bonded in the helix state. This is as expected since only the terminal C=O can possibly hydrogen bond. In the coil state, on average about 2.5 C=O groups are hydrogen bonded. As a fraction of the total number of  guanidininum groups in the inner shell,  less than 17\% of the available groups can be considered as hydrogen bonding with the solute in the coil state. This observation appears to be consistent with experiments that show that hydrogen bonding does not explain denaturation by GdmCl \cite{englander:gdmcl09}. 

The h-bonding analysis together with the trend for the $Q=0$ peptide suggests the importance of non-specific (promiscuous) van~der~Waals interaction in the interaction of GdmCl with peptides. But as seen below, including the packing contribution reverses the contributions from favorable local protein-cosolute interactions. Hints of the same reversal behavior can be seen by comparing the revised chemistry contribution (Table~\ref{tb:all}), but we do not show this for brevity.

\subsection{Local chemistry plus packing contribution}
\begin{figure*}[h!]
\includegraphics[scale=1]{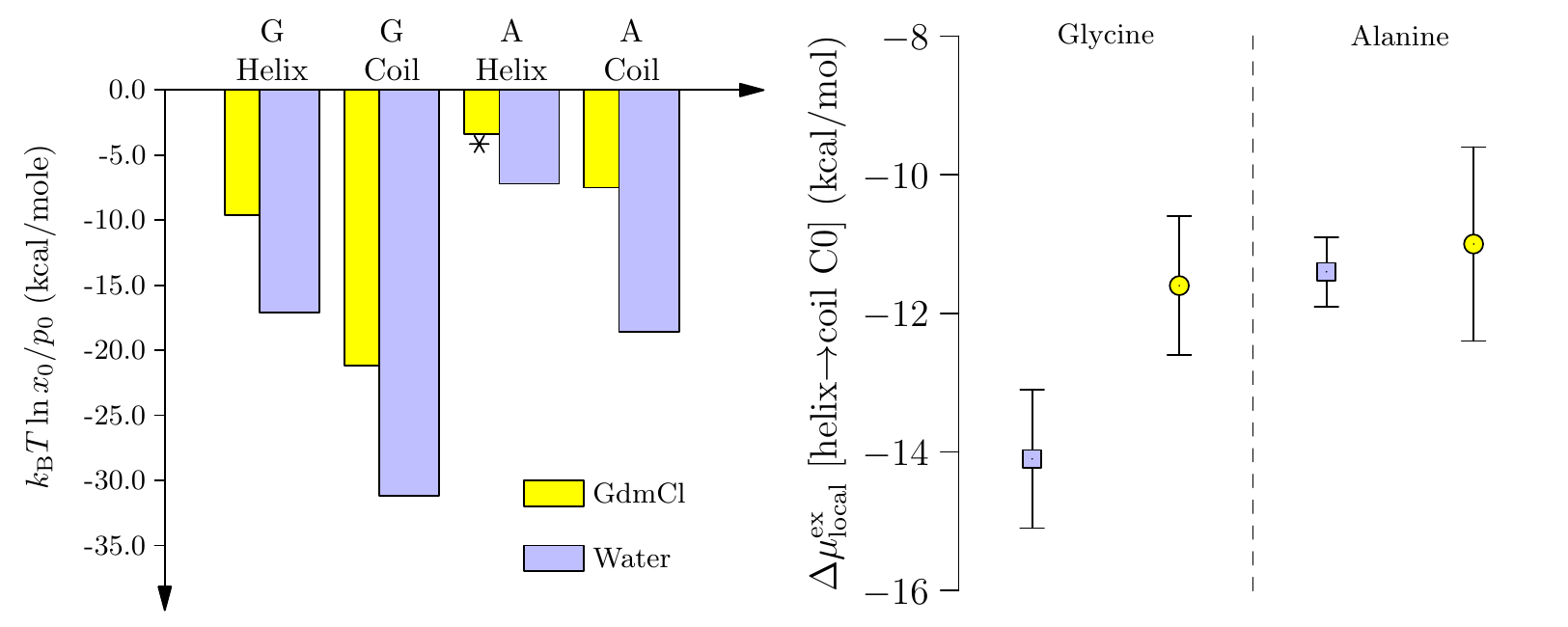}
\caption{\underline{Left panel}: SE packing plus revised chemistry contribution to the solvation free energy. The asterisk for Ala-helix in aqueous GdmCl indicates that the negative of the actual value is shown to simplify the graphing. \underline{Right panel}: Local contribution to the solvation free energy change in the helix-to-coil C$_0$ transition in water (filled blue squares) and in aqueous GdmCl (filled yellow circle). Note that these are not free energies for transfer across solvents, and hence the color code is different from those in Figs.~\ref{fg:dmu},~\ref{fg:packing}, and~\ref{fg:Gpacking}. Standard error of the mean is shown at $1\sigma$.}\label{fg:local}
\end{figure*}

Figure~\ref{fg:local} (left panel) shows the net local contribution to the solvation free energy, i.e.\ the SE packing contribution plus the revised chemistry contribution, which is the same as packing contribution for the $\Sigma_{G}$ envelope (Fig.~\ref{fg:Gpacking}) and the total chemistry contribution (Fig.~\ref{fg:chemistry}). 
Except for the solvation of the alanine helix in aqueous GdmCl, a case where packing dominates the chemistry contribution, the short-range
contribution to solvation is negative, and more so for the aqueous GdmCl solution. This suggests that despite the larger packing contribution in aqueous GdmCl (Fig.~\ref{fg:packing}), the local attractive contributions are favorable enough to render the net local contribution favorable. However, as Fig.~\ref{fg:local} (right panel) shows, relative to water, 
the favorable free energy change for unfolding the peptide in aqueous GdmCl is less than that in water. Thus, for the systems studied here, local attractive contributions, i.e.\ solute interaction with solvent components in the first solvation shell, cannot explain the expected action of GdmCl in favoring the
extended conformation of the peptide (Fig.~\ref{fg:dmu}). 

\begin{figure*}[h!]
\includegraphics[scale=1]{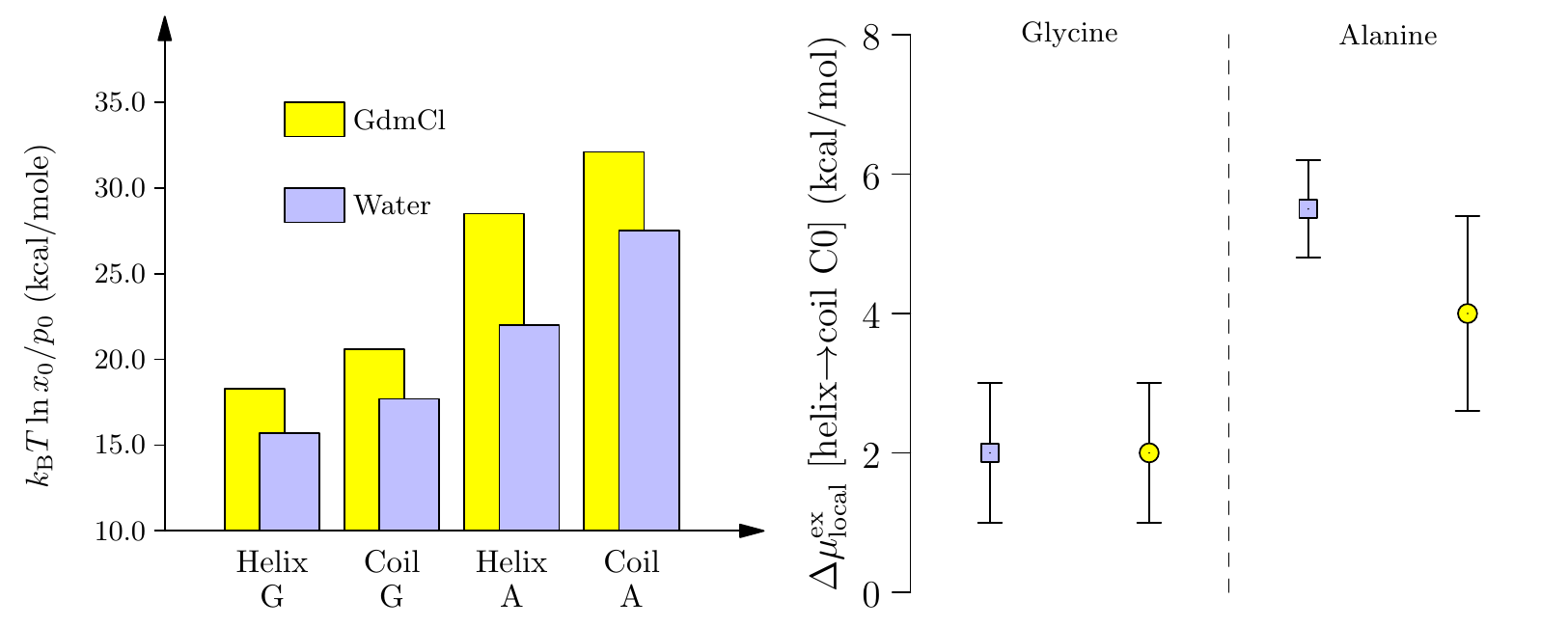}
\caption{Results for the $Q=0$ analogues of the peptides. Rest as in Fig.~\ref{fg:local}.}\label{fg:localneut}
\end{figure*}

For a peptide lacking partial charges, the change in the local contribution for the helix-to-coil C$_0$ transition is comparable for the glycine model (Fig.~\ref{fg:localneut}). Interestingly, for the alanine peptide, a system that ought to have more favorable van~der~Waals interactions with the medium, the change in the local contribution is favored in aqueous GdmCl relative to that in water, suggesting that dispersion interactions play an important role in the solvation thermodynamics. 

\subsection{Long-range contributions}

Table~\ref{tb:outer}  collects the long-range contribution on the basis of $P^{(0)}(\varepsilon|\lambda_G)$ (Eq.~\ref{eq:for}). 
As already noted, the conditional binding energy distribution is Gaussian to an excellent approximation \cite{tomar:bj2013,tomar:jpcb14,tomar:jpcb16,asthagiri:gly17}.  The binding energy $\varepsilon$ is a sum of van~der~Waals, $\varepsilon_{vdW}$,  and electrostatic, $\varepsilon_{elec}$, contributions. Here we ask if $P^{(0)}(\varepsilon|\lambda_G) = P^{(0)}(\varepsilon_{vdW}|\lambda_G) \cdot P^{(0)}(\varepsilon_{elec}|\lambda_G)$, i.e.\ whether the long-range van~der~Waals and electrostatic contributions are uncorrelated, and if each of the component distributions are also Gaussian. Table~\ref{tb:outer} shows that this is indeed the case within statistical uncertainties of the calculation, reaffirming the validity of a Gaussian model and our demarcation of inner- and outer-shells. (We emphasize that all the specific interactions are part of the inner-shell, local contribution and
all the non-specific, uncorrelated contributions are part of the outer-shell, long-range contribution.)
\begin{table*}[h!]
\caption{Long-range contribution to the chemical potential for the deca-alanine peptide. $\mu^{\rm ex} [P^{(0)}(\varepsilon|\lambda_G)]$ (Eq.~\ref{eq:for}) is obtained using a Gaussian model. The electrostatic, $\mu^{\rm ex}_{\rm elec}$, and van~der~Waals, $\mu^{\rm ex}_{vdW}$, contributions are obtained using a Gaussian description of the distribution of the respective interaction energies. Standard error of the mean on the net contribution is at the $1\sigma$ level. Standard errors for the component values in water have been ignored as it proves inconsequential in the $\Delta$ values. All values are in kcal/mol.}\label{tb:outer}
\begin{tabular}{c c r r r r}
Conformer & Solvent           & $\mu^{\rm ex}_{\rm lr}$ & $\mu^{\rm ex}_{vdW}$ & $\mu^{\rm ex}_{elec}$ &  $\mu^{\rm ex}_{vdW} + \mu^{\rm ex}_{elec}$ \\ \hline
\textbf{Helix} & Water              &     $-31.6\pm0.2$                          &    $-18.1$               &  $-13.3$             &    $-31.4$ \\
                      & GdmCl(aq)     &     $-35.6\pm0.6$                           &    $-23.3\pm0.2$   &  $-12.5\pm0.1$  &    $-35.8\pm0.2$ \\
                      & $\Delta$         &      $-4.0\pm0.6$                            &    $-5.2\pm0.2$     &  $0.8\pm0.1$     &    $-4.4\pm0.2$ \\ \hline
\textbf{Coil} & Water             &      $-27.6\pm0.1$                          &  $-24.7$                        & $-2.9$                         & $-27.6$ \\
                   & GdmCl(aq)   &       $-35.2\pm 0.5$                         & $-32.6\pm0.5$                         & $-2.7\pm0.1$                          & $-35.3\pm0.5$ \\ 
                   & $\Delta$       &       $-7.6\pm0.5$                            & $-7.9\pm0.5$                            & $0.2\pm0.1$                           & $-7.7\pm0.5$ \\ \hline
\end{tabular}
\end{table*}

\begin{figure*}[h!]
\includegraphics[scale=0.9]{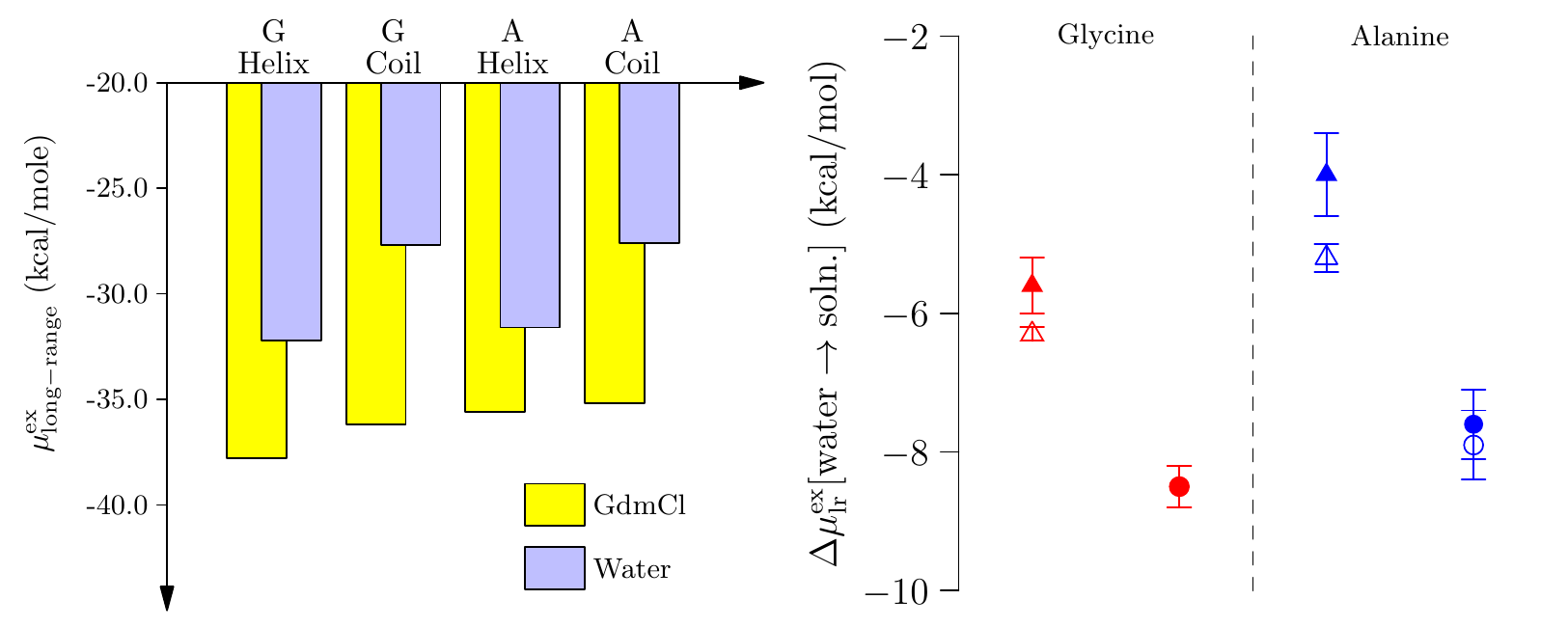}
\caption{\underline{Left panel}: Long-range contribution to the solvation of the physical peptide. \underline{Right panel}: Long-range contributions to the
water-to-GdmCl(aq) transfer free energy for the physical peptides (filled symbols) and its $Q=0$ analogue (open symbols). For the $Q=0$ peptide the
 long-range contributions are given in Table~\ref{tb:all} in the Appendix. $\triangle$: helix; $\bigcirc$: $C_0$ coil conformer. Standard error of the mean is shown at $1\sigma$.}\label{fg:outer}
\end{figure*}

Figure~\ref{fg:outer} shows the long-range contribution to solvation for both the deca-glycine and deca-alanine peptides. The long-range contribution in the water-to-GdmCl(aq) transfer free energy is more negative for the coil state than the helix state. Further, since the net local contribution disfavors the helix-to-coil transition in GdmCl(aq) relative to that in water (Fig.~\ref{fg:local}), it is clear that the long-range contributions are the deciding factor in the net transfer free energy change (Fig.~\ref{fg:dmu}). 

The long-range contributions to hydration always favor the helix relative to the coil. This is ultimately due to the favorable electrostatic interaction between the helix macro-dipole and the surrounding solvent \cite{tomar:jpcb16}. The favorable solvation of the helix macro-dipole also holds in GdmCl(aq), but now in addition the van~der~Waals interactions also play an important role. Table~\ref{tb:outer} shows that electrostatic contributions nearly cancel in the water-to-GdmCl(aq) transfer free energy, but long-range solute-solvent van~der~Waals interactions in GdmCl(aq) are always more favorable than in water. It is this 
contribution --- the sum total of weak, but attractive van~der~Waals interactions --- that is decisive in recovering the observed free energy (Fig.~\ref{fg:dmu}). 

\section{Discussions}

As seen above, in the action of the cosolute, turning-on peptide-solvent electrostatic interactions attenuates the effect of the cosolute. There are two reasons for this. First, numerically there are many more water molecules in the system and thus in the inner-shell of the peptide, and second, there are more possible ways for water molecules to interact with the peptide than is the case for the Gdm$^+$ cation. Thus, in the presence of electrostatic interactions there will greater cohesion between the peptide and water \cite{tomar:jpcb16,asthagiri:gly17} and more water molecules will be present in the inner shell inhibiting the interaction of the cosolute. But van~der~Waals interactions are less sensitive to the orientation of the cosolute and hence emerges as a dominant factor in the mode of action of GdmCl. 

There are also important limitations in the present study, and we address these next. 

Our studies above have focused on two limiting structures of the peptide, the compact helical conformation and the least solvated extended conformation of the peptide, labelled $C_0$ state in our previous study \cite{tomar:jpcb16}. As already noted above, from Eq.~\ref{eq:multistate} we can conclude that the free energy of the unfolded state will be lower than the free energy of the $C_0$ state. Exploratory adaptive bias force \cite{abf1,abf2} calculations of the PMF along the end-to-end distance of the unfolded peptide shows that the free energy surface is biased towards the extended conformations for GdmCl than for water, consistent with the role of GdmCl in favoring the unfolded structure of proteins. We can expect that including more of the extended conformations with the appropriate conformational weights (Eq.~\ref{eq:multistate}) would predict a lower free energy of the coil state in aqueous GdmCl. 
% That is, with better sampling we expect the free energy change for the helix-to-coil transition to be more favorable in GdmCl than in water. 
Unfortunately, given the demands of the free energy calculations, we have not pursued this path here.  But, as already discussed, the analysis of solvation free energies of the two limiting structures is sufficient to support the central insights drawn in this work. 

The second outstanding issue deals with relating our results with those based on the preferential interaction approach. It is often implicitly assumed that the water-to-cosolute transfer free energy of the solute  is favored in solutions where the cosolute preferentially interacts with the solute.  But, as has been pointedly noted before  \cite{timasheff:93rev}, knowledge of the preferential interaction parameter at a single solvent composition gives no information about the transfer free energy. Indeed it is possible for the preferential interaction to be positive (i.e.\ favorable) and the transfer free energy to also be positive (i.e.\ unfavorable) \cite{timasheff:93rev}. To relate preferential interaction to transfer free energies one needs the composition dependence of the preferential interaction parameter together with the activity coefficient of the cosolute solution across the composition range. Obtaining these quantities with sufficient statistical resolution is a demanding task and we must necessarily leave this for future studies.

\section{Conclusions}

Our previous studies led to the finding that peptide-solvent attractive interactions overcome the hydrophobic contribution to favor the unfolded state of the protein. The present study builds on those efforts and examines the role of GdmCl in modulating the solvophobic and solvophilic contributions. Aqueous GdmCl increases the solvophobic contribution, but the solvophilic contribution is also amplified. The local solvophilic contributions temper the solvophobic contributions, but are not sufficient to predict the net free energy change. We also find that hydrogen bonding is not consequential in the local interaction of Gdm+ with the peptide. We find that the contribution to the free energy from the long-range interaction of the solute with solvent components outside the first solvation shell is decisive in recovering the net transfer free energy change. In the transfer free energy, the long-range contribution is primarily derived from weak, but attractive, dispersion interactions. The latter finding is important because such long-range contributions are usually ignored in attempts to rationalize osmolyte or denaturant effect in terms of local binding of the cosolute with the protein.

\section*{Acknowledgments}
We gratefully acknowledge computing support from the National Energy Research Scientific Computing Center (NERSC), a DOE Office of Science User Facility supported by the Office of Science of the U.S. Department of Energy under Contract No. DE-AC02-05CH11231. 

\section{Appendix}
Table~\ref{tb:all} below collects the component free energy values used in the analysis in the main text. Results obtained using our adaptation of the Kirkwood-Buff-based forcefield model for GdmCl \cite{smith:gdcl04} are similar to the ones noted below for the CGenFF forcefield and hence are not shown. 
\begin{table*}
\caption{Free energy components (Eqs.~\ref{eq:qc} and~\ref{eq:qc1}) in the solvation of the peptide in the specified conformer in the given solvent. 
Standard error of the mean is at the $1\sigma$ level. All values are in kcal/mole.}\label{tb:all}
\begin{tabular}{c c r r r r r} \hline 
Conformer & Solvent          & $\mu^{\rm ex}_{\rm pack} (\lambda_{SE}) $ & $\mu^{\rm ex}_{\rm pack} (\lambda_{G})$ & $\mu^{\rm ex}_{\rm chem} (\lambda_{G})$ & $\mu^{\rm ex}_{\rm lr}$ & $\mu^{\rm ex}_{\rm net}$ \\ \hline
\textbf{Ala-helix} &  Water               &     $44.3\pm0.4$                                           &  $87.4\pm0.4$ & $-94.6\pm0.1$ & $-31.6\pm0.2$ & $-38.8\pm 0.5$ \\
                            & GdmCl(aq)       &     $57.9\pm1.0$    & $109.2\pm 1.0$ & $-105.8\pm 0.5$ & $-35.6 \pm 0.6$ & $-32.2\pm 1.0$ \\  
                            &  $\Delta$          &     $13.6\pm1.0$ & $21.8\pm 1.0$ & $-11.2\pm0.5$ & $-4.0\pm0.6$ & $6.6\pm 1.0$ \\ 

\textbf{Ala-coil} &  Water               &     $58.4\pm0.2$ & $127.8\pm0.3$ & $-146.4\pm0.2$ & $-27.6\pm0.1$ & $-46.2\pm0.4$ \\
                          &  GdmCl(aq)       &      $67.7\pm0.4$ & $151.4\pm0.6$ & $-158.9\pm0.5$ & $-35.2\pm0.5$ & $-42.7\pm0.9$ \\ 
                          & $\Delta$            &       $9.3\pm0.5$ & $23.6\pm0.7$ & $-12.5\pm0.5$ & $-7.6\pm0.5$ & $3.5\pm1.0$ \\ 

\textbf{Ala-helix $Q=0$} &  Water  &     $44.3\pm0.4$  &  $87.4\pm0.4$ & $-65.4\pm0.3$ & $-18.1\pm0.0$ & $3.9\pm0.5$ \\
                            & GdmCl(aq)       &     $57.9\pm1.0$  & $109.2\pm 1.0$ & $-80.7\pm 0.5$ & $-23.3\pm0.2$ & $5.2\pm 1.0$ \\  
                            &  $\Delta$          &     $13.6\pm1.0$ & $21.8\pm 1.0$ & $-15.3\pm0.6$ & $-5.2\pm0.2$ & $1.3\pm 1.0$ \\                           

\textbf{Ala-coil $Q=0$} &  Water    &     $58.4\pm0.2$ & $127.8\pm0.3$ & $-100.3\pm0.4$ & $-24.7\pm0.0$ & $2.8\pm0.5$ \\
                          &  GdmCl(aq)       &      $67.7\pm0.4$ & $151.4\pm0.6$ & $-119.3\pm0.5$ & $-32.6\pm0.5$ & $-0.5\pm0.9$ \\ 
                          & $\Delta$            &       $9.3\pm0.5$ & $23.6\pm0.7$ & $-19.0\pm0.6$ & $-7.9\pm0.5$ & $-3.3\pm1.0$ \\ 

\textbf{Gly-helix} &  Water               &     $37.9\pm0.6$ &  $78.0\pm0.7$ & $-95.1\pm0.3$ & $-32.2\pm0.3$ & $-49.3\pm 0.8$ \\
                            & GdmCl(aq)       &     $46.6\pm0.5$  & $94.0\pm0.6$ & $-103.6\pm 0.4$ & $-37.8 \pm 0.2$ & $-47.4 \pm 0.7$ \\  
                            &  $\Delta$          &     $8.7\pm0.8$ & $16.0\pm0.9$ & $-8.5\pm0.5$ & $-5.6\pm0.4$ & $1.9\pm 1.0$ \\ 

\textbf{Gly-coil} &  Water                &    $48.7\pm0.3$ & $115.9\pm0.5$ & $-147.1\pm0.3$ & $-27.7\pm0.1$ & $-58.9\pm0.6$ \\
                            & GdmCl(aq)       &  $58.0\pm0.3$ & $139.0\pm0.6$ & $-160.2\pm0.5$ & $-36.2 \pm0.3$ & $-57.4\pm0.8$ \\
                            &  $\Delta$          &    $9.3\pm0.4$ & $23.1\pm0.8$ & $-13.1\pm0.6$ & $-8.5\pm0.3$ & $1.5\pm1.0$ \\ 

\textbf{Gly-helix $Q=0$} &  Water  &     $37.9\pm0.6$ &  $78.0\pm0.7$ & $-62.3\pm0.3$ & $-17.6\pm0.0$ & $-1.9\pm 0.8$ \\
                            & GdmCl(aq)       &     $46.6\pm0.5$  & $94.0\pm0.6$ & $-75.7\pm 0.4$ & $-23.9 \pm 0.1$ & $-5.6 \pm 0.7$ \\  
                            &  $\Delta$          &     $8.7\pm0.8$ & $16.0\pm0.9$ & $-13.4\pm0.5$ & $-6.3\pm0.1$ & $-3.7\pm 1.0$ \\
                            
\textbf{Gly-coil $Q=0$}  &  Water    &    $48.7\pm0.3$ & $115.9\pm0.5$ & $-98.2 \pm0.3$ & $-24.1\pm0.0$ & $-6.4\pm0.7$ \\
                            & GdmCl(aq)       &  $58.0\pm0.3$ & $139.0\pm0.6$ & $-118.4\pm0.4$ & $-32.6 \pm0.3$ & $-12.0\pm0.8$ \\
                            &  $\Delta$          &    $9.3\pm0.4$ & $23.1\pm0.8$ & $-20.2\pm0.5$ & $-8.5\pm0.3$ & $-5.6\pm 0.9$ \\ \hline
\end{tabular}
\end{table*}

\newpage
% \bibliography{refs}

\end{document}